\documentclass[twocolumn,prb,groupedaddress,amsmath,amssymb,floatfix]{revtex4-1}
\usepackage{graphicx}
\usepackage{amssymb}
\usepackage{dcolumn}
\usepackage{bm}
\def\dps{\displaystyle}
\def\m#1{\mathrm{#1}}
\def\Eq#1{(\ref{eq:#1})}
\def\d{\mathrm{d}}

\def\epsilon{\varepsilon}
\def\theta{\vartheta}
\def\rho{\varrho}

\def\diag{\mathop\mathrm{diag}}

\begin{document}

\title{The structure of fluids with impurities}

\author{M. Bier and L. Harnau$^*$}

\email{harnau@fluids.mpi-stuttgart.mpg.de}

\affiliation{
Max-Planck-Institut f\"ur Intelligente Systeme, Heisenbergstr.~3, 
70569 Stuttgart, Germany,
\\and Institut f\"ur Theoretische und Angewandte Physik, Universit\"at Stuttgart,
Pfaffenwaldring 57, 
70569 Stuttgart, Germany}
\date{\today}

\begin{abstract}
The influence of dilute impurities on the structure of a fluid solvent is investigated 
theoretically. General arguments, which do not rely on particular models, are used to 
derive an extension of the Ornstein-Zernike form for the solvent structure factor at small 
scattering vectors. It is shown that dilute impurities can influence the solvent structure 
only if they are composed of ions with significantly different sizes.
Non-ionic impurities or ions of similar size are shown to not alter the
solvent structure qualitatively. This picture is compatible with available 
experimental data. The derived form of the solvent structure factor is expected to be useful 
to infer information on the impurity-solvent interactions form measured scattering intensities.  
\end{abstract}
\maketitle

{\it Dedicated to Matthias Ballauff on the occasion of his 60$^{th}$ birthday.}

%===================================================================================================

\section{Introduction}

Scattering techniques are powerful approaches to infer structural properties of soft materials.
In various combined experimental and theoretical studies, Matthias Ballauff has contributed much 
insight into this topic by successfully comparing partial structure factors calculated from 
Ornstein-Zernike equations with experimentally determined scattering data of dissolved dendrimers 
\cite{liko:02,rose:06,harn:07,rose:09}, bottle-brush polymers \cite{boli:07,boli:09},
polyelectrolyte brushes \cite{henz:08,henz:11}, plate-like colloids \cite{li:05}, and 
nanoparticles \cite{webe:07,roch:11}. In these earlier studies the Ornstein-Zernike equations 
have been employed in the study of pair correlations of the dissolved particles. 
However, due to complex preparation procedures, many fluids contain impurities, which potentially
influence the solvent structure factor even in the case of highly diluted impurities
\cite{Sadakane2006,Sadakane2007a,Sadakane2007b,Sadakane2011}.
In the following we address the problem of how impurities can alter the form of the  
solvent structure factor. By means of general, model independent arguments based on Ornstein-Zernike 
equations we derive a general asymptotic form of the solvent structure factor.
This form of the solvent structure factor and the insight gained from it is expected
to be useful also for the analysis of future studies involving impurities in fluid systems.

In the next Sec.~\ref{Sec:math} the setting is defined and a detailed mathematical derivation of
the asymptotic structure factor is given.
The consequences of this asymptotic form are discussed in Sec.~\ref{Sec:discuss}.
Finally, conclusions and a summary are given in Sec.~\ref{Sec:conclude}.

%===================================================================================================

\section{\label{Sec:math}Mathematical part}

%---------------------------------------------------------------------------------------------------

\subsection{Setting}

Consider a spatially uniform fluid mixture of three components of which component 1 is referred
to as the ``solvent'' whereas components 2 and 3 are called ``impurities''.
The interaction potentials $U_{ij}(r)$ between two particles of components $i$ and $j$ are assumed
to be isotropic and vanish\-ing at infinite particle separation ($r\to\infty$).
Note that no particular asymptotic decay of $U_{ij}(r\to\infty)$ is assumed.
The mole fraction $x_i$ of component $i\in\{1,2,3\}$ may be expressed in terms of the mole fraction
$x = x_2+x_3$ of impurities and the composition $\phi\in[0,1]$ of the impurities such that 
$x_1 = 1-x$, $x_2=\phi x$ and $x_3=(1-\phi)x$.
For \emph{ionic} impurities with valencies $z_2,z_3>0$ of the respective components the constraint 
$z_2x_2 = z_3x_3$ of local charge neutrality of the bulk fluid leads to the composition
$\phi = z_3/(z_2+z_3)$, whereas the composition $\phi$ of \emph{non-ionic} impurities is not 
restricted to a particular value.
 
%---------------------------------------------------------------------------------------------------

\subsection{\label{Subsec:strucfac}Solvent structure factor}

In general the partial structure factors $S_{ij}(q)$ with $i,j\in\{1,2,3\}$ of a fluid with total
number density $\rho$ are given by \cite{Hansen1986}
\begin{equation}
   S_{ij}(q) 
   = 
   \sqrt{x_ix_j}\delta_{ij} + x_ix_j\rho\widehat{h}_{ij}(q),
   \label{eq:Sij}
\end{equation}
where 
\begin{equation}
   \widehat{h}_{ij}(q) 
   = 
   \frac{4\pi}{q}\int\limits_0^\infty\d r\ rh_{ij}(r)\sin(qr)
   \label{eq:hij}
\end{equation}
is the (three-dimensional) Fourier transform of the total correlation function $h_{ij}(r) =
g_{ij}(r)-1$. Here $q$ denotes the magnitude of the scattering vector and $g_{ij}(r)$ is the 
radial distribution function. As the name implies, the total correlation function $h_{ij}(r)$ 
gives a complete information about all the correlations between a pair of particles of components 
$i$ and $j$. By considering $S_{ij}(q)$ and $\widehat{h}_{ij}(q)$ as the $(i,j)$-components
of the matrices $\underline{S}(q)$ and $\underline{\widehat{h}}(q)$, respectively, and with the
diagonal matrix $\underline{x}:=\diag(x_1,x_2,x_3)$ one can rewrite Eq.~\Eq{Sij} in the matrix
form
\begin{equation}
   \underline{S}(q) 
   = 
   \sqrt{\underline{x}}
   \big(\underline{1}+\rho\sqrt{\underline{x}}\underline{\widehat{h}}(q)\sqrt{\underline{x}}\big)
   \sqrt{\underline{x}}.
   \label{eq:Sh}
\end{equation}

The Ornstein-Zernike equation links $\widehat{h}_{ij}(q)$ to the Fourier transform 
$\widehat{c}_{ij}(q)$ of the direct correlation function $c_{ij}(r)$ \cite{Hansen1986}:
\begin{equation}
   \widehat{h}_{ij}(q) = \widehat{c}_{ij}(q) + \rho\sum_kx_k\widehat{c}_{ik}(q)\widehat{h}_{kj}(q)
   \label{eq:OZij}
\end{equation}
or in an equivalent matrix form
\begin{equation}
   \big(\underline{1}+\rho\sqrt{\underline{x}}\underline{\widehat{h}}(q)\sqrt{\underline{x}}\big)
   \big(\underline{1}-\rho\sqrt{\underline{x}}\underline{\widehat{c}}(q)\sqrt{\underline{x}}\big)
   =
   \underline{1}.
   \label{eq:OZ}
\end{equation}
Hence the matrix of partial structure factors $\underline{S}(q)$ is given by
\begin{equation}
   \underline{S}(q)
   =
   \sqrt{\underline{x}}
   \big(\underline{1}-\rho\sqrt{\underline{x}}\underline{\widehat{c}}(q)\sqrt{\underline{x}}\big)
   ^{-1}
   \sqrt{\underline{x}}.
   \label{eq:Sc}
\end{equation}

Introducing $\gamma_{ij} := \rho\sqrt{x_ix_j}\widehat{c}_{ij}(q)$ for given $q$ one obtains
from Eq.~\Eq{Sc} by means of Cramer's rule the solvent partial structure factor
\begin{eqnarray}
\lefteqn{S_{11}(q) = } \nonumber
\\&&
\frac{x_1}{\dps 1-\gamma_{11} - \frac{\gamma_{12}^2(1-\gamma_{33}) + 
                                         \gamma_{13}^2(1-\gamma_{22}) + 
                                         2\gamma_{12}\gamma_{13}\gamma_{23}}
                                        {(1-\gamma_{22})(1-\gamma_{33}) - \gamma_{23}^2}}.
   \label{eq:S11}
\end{eqnarray}

%---------------------------------------------------------------------------------------------------

\subsection{\label{Subsec:dcf}Direct correlation functions}

Since the Ornstein-Zernike equation is a definition of the direct correlation function $c_{ij}(r)$, 
it does not yield a method of calculating this correlation function.
In order to proceed, the asymptotic behaviour $\widehat{c}_{ij}(q\to0)$ of the Fourier transform
of the direct correlation function $c_{ij}(r)$ is required.
It is known \cite{Hansen1986} that $c_{ij}(r>R_{ij})\simeq-\beta U_{ij}(r)$ with $1/\beta=k_BT$ for
sufficiently large $R_{ij}>0$.
Hence $\widehat{c}_{ij}(q) = \widehat{c}_{ij}^{<}(q) + \widehat{c}_{ij}^{>}(q)$ with
\begin{equation}
   \widehat{c}_{ij}^{<}(q) 
   =
   \frac{4\pi}{q}\int\limits_0^{R_{ij}}\d r\ rc_{ij}(r)\sin(qr)
   \label{eq:cless}
\end{equation}
being an even entire function of $q$ and
\begin{equation}
   \widehat{c}_{ij}^{>}(q) 
   =
   -\frac{4\pi}{q}\int\limits_{R_{ij}}^\infty\d r\ r\beta U_{ij}(r)\sin(qr).
   \label{eq:cgtr}
\end{equation}

If $\beta U_{ij}(r) = A_{ij}/r$ one obtains 
\begin{equation}
   \widehat{c}_{ij}^{>}(q) \simeq -\frac{4\pi A_{ij}}{q^2} + f_{ij}^{>}(q)
   \label{eq:cgtrCoulomb}
\end{equation}
with an even entire function $f_{ij}^{>}(q)$ of $q$.
However, if $\beta U_{ij}(r)$ decays faster than $\mathcal{O}(1/r)$, including power laws
$\propto 1/r^n, n \geq 2,$ and all short-ranged interaction potentials, $\widehat{c}_{ij}^{>}(q)$
is an even entire function of $q$. 
Hence the Fourier transform $\widehat{c}_{ij}(q)$ is of the general form 
\begin{equation}
   \widehat{c}_{ij}(q) \simeq -\frac{4\pi A_{ij}}{q^2} + f_{ij}(q)
   \label{eq:cgeneral}
\end{equation}
with an even entire function $f_{ij}(q)$ of $q$ and with a coefficient $A_{ij}$,
which vanishes if $\beta U_{ij}(r\to\infty)$ decays faster than $\mathcal{O}(1/r)$.

%---------------------------------------------------------------------------------------------------

\subsection{\label{Subsec:asymps11}Asymptotic solvent structure factor} 

For small mole fraction $x\ll1$ of impurities and inverse scattering vector magnitudes $1/q$ larger 
than the particle sizes, the general form Eq.~\Eq{cgeneral} of the Fourier transform $\widehat{c}_{ij}(q)$ 
of the direct correlation function $c_{ij}(r)$ leads to the following expressions of $\gamma_{ij}$ 
defined in Subsec.~\ref{Subsec:strucfac}:
\begin{align}
   \gamma_{11} & \simeq \rho(c_{11}^{(0)} + c_{11}^{(1)}q^2) \label{eq:gamma11}\\
   \gamma_{12} & \simeq \rho\sqrt{x\phi}(c_{12}^{(0)} + c_{12}^{(1)}q^2) \\
   \gamma_{13} & \simeq \rho\sqrt{x(1-\phi)}(c_{13}^{(0)} + c_{13}^{(1)}q^2) \\
   \gamma_{22} & \simeq \rho x\phi\Big(-\frac{4\pi\ell_Bz_2^2}{q^2} + 
                             c_{22}^{(0)} + c_{22}^{(1)}q^2\Big)  \label{eq:gamma22}\\
   \gamma_{33} & \simeq \rho x(1-\phi)\Big(-\frac{4\pi\ell_Bz_3^2}{q^2} + 
                             c_{33}^{(0)} + c_{33}^{(1)}q^2\Big) \\
   \gamma_{23} & \simeq \rho x\sqrt{\phi(1-\phi)}\Big(\frac{4\pi\ell_Bz_2z_3}{q^2} + 
                             c_{23}^{(0)} + c_{23}^{(1)}q^2\Big),\label{eq:gamma23}
\end{align}
where $c_{ij}^{(0)}$ and $c_{ij}^{(1)}$ are the expansion coefficients of the entire parts 
$f_{ij}(q)$ of $\widehat{c}_{ij}(q)$ (see Eq.~\Eq{cgeneral}) in $q=0$ and $A_{22}=l_B z_2^2$, 
$A_{33}=l_B z_3^2$, $A_{23}=-l_B z_2 z_3$, where $\ell_B$ denotes the Bjerrum length.
In the case of non-ionic impurities, i.e., $z_2=z_3=0$, no pole occurs at $q=0$ in 
Eqs.~\Eq{gamma22}--\Eq{gamma23}. 

In the absence of impurities, i.e., for $x=0$, one obtains the well-known Ornstein-Zernike 
structure factor $S_{11}(q)=(1-\rho(c_{11}^{(0)} + c_{11}^{(1)}q^2))^{-1}$ from Eqs.~\Eq{S11} and
\Eq{gamma11}.
From the relation $S_{11}(0)>0$, which holds due to the fact that $S_{11}(0)$ is the ratio between
the compressibility of the solvent and that of a gas of non-interacting particles of the same 
temperature and density  \cite{Hansen1986}, one infers $1-\rho c_{11}^{(0)}\geq0$.
Moreover, $c_{11}^{(1)}<0$ is required if the bulk phase diagram of the pure solvent exhibits critical 
points and if the pure solvent is uniform throughout the whole one-phase region of the bulk phase 
diagram, i.e., $S_{11}(q)$ for $q>0$ does not diverge even close to critical points.
Note that these conditions are not fulfilled by a fluid of hard spheres, because there is no
critical point in the corresponding bulk phase diagram, which is independent of temperature and 
exhibits only a first-order phase transition.

Inserting Eqs.~\Eq{gamma11}--\Eq{gamma23} into Eq.~\Eq{S11} and considering only the dominant 
contributions in the mole fraction $x\ll1$ of impurities leads to the asymptotic form
\begin{equation}
   S_{11}(q) \simeq 
   \frac{S_{11}(0)}{\dps 1 + (\xi q)^2\left(1 - \frac{g^2}{1 + (q/\kappa)^2}\right)},
   \label{eq:S11asym}
\end{equation}   
where 
\begin{equation}
   S_{11}(0) = \frac{1}{1-\rho(1-x)c_{11}^{(0)}-\rho^2x(\phi c_{12}^{(0)}+(1-\phi) c_{13}^{(0)})^2}
\end{equation}
is related to the compressibility of the solvent \cite{Hansen1986}, 
$\xi = \sqrt{-\rho c_{11}^{(1)}S_{11}(0)}$ is the bulk correlation length and
$\kappa = \sqrt{8\pi\ell_BI}$ is the inverse Debye length with 
$I=\rho x(z_2^2\phi+z_3^2(1-\phi))/2$ denoting the ionic strength. The quantity
\begin{equation}
   g^2 := \frac{(c_{12}^{(0)} - c_{13}^{(0)})^2}{-4\pi\ell_Bc_{11}^{(1)}(z_2 + z_3)^2}
   \label{eq:g2}
\end{equation}
measures the contrast of impurity-solvent interactions. Hence the normalized solvent structure 
factor $S_{11}(q)/S_{11}(0)$ depends on the two dimensionless parameters $\kappa \xi$ and $g^2$ 
which are related to material parameters such as $\rho, x, \phi, z_2, z_3$, and $l_B$.
The form of Eq.~\Eq{S11asym} has previously been derived in Ref.~\cite{Onuki2004} within a specific
model of electrolyte solutions.
The derivation presented here demonstrates that the asymptotic form Eq.~\Eq{S11asym} of the solvent 
structure factor $S_{11}(q)$ is independent of any specific model.
The consequences of Eqs.~\Eq{S11asym}--\Eq{g2} and the physical meaning of the quantity $g^2$ are 
discussed in the following section. 

%===================================================================================================

\section{\label{Sec:discuss}Discussion}

%---------------------------------------------------------------------------------------------------

\subsection{\label{Subsec:forms}Forms of the solvent structure factor}

Considering $S_{11}(q)$ in Eq.~\Eq{S11asym} one immediately recognizes 
\begin{equation}
   S_{11}(q)\simeq
   \left\{\begin{array}{ll}
      \dps \frac{S_{11}(0)}{1-g^2(\kappa\xi)^2 + (\xi q)^2} & ,\,\, q\gg\kappa \\[15pt]
      \dps \frac{S_{11}(0)}{1+(1-g^2)(\xi q)^2}             & ,\,\, q\ll\kappa.
   \end{array}\right.  
\end{equation}
In particular, for non-ionic impurities, i.e., for $\kappa=0$, the solvent structure factor
$S_{11}(q)\simeq S_{11}(0)/(1+(\xi q)^2)$ is not altered to leading order in the mole fraction $x$.
However, ionic impurities due to, e.g., alkali halides of ionic strength 
$I\approx1\,\m{mM}$ in water at room temperature, i.e., $x\approx 4\cdot10^{-5}$, lead to 
a Debye length $1/\kappa \approx 10\,\m{nm}$ which is much larger than the particle size.

\begin{figure}[ht!]
\includegraphics[width=8.5cm]{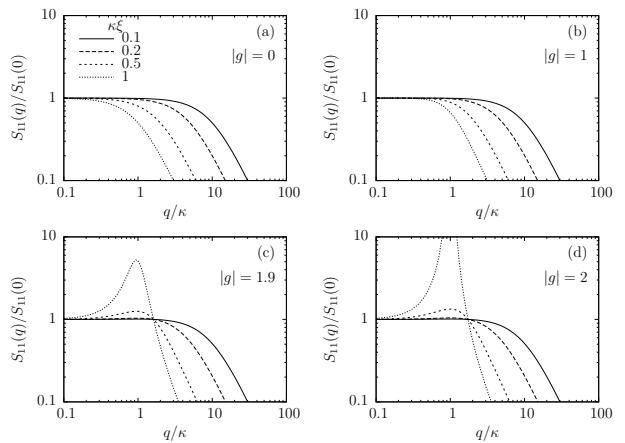}
\caption{Influence of dilute ionic impurities on the solvent structure factor $S_{11}(q)$.
The bulk correlation length is denoted by $\xi$, whereas the inverse Debye length
is given by $\kappa$. The contrast of impurity-solvent interactions is measured by the 
parameter $|g|$ (see Eq.~\Eq{g2}). For $|g|\leq1$ (panels (a) and (b)) a monotonic structure 
factor is found, whereas a non-monotonic behaviour with a maximum close to the Debye length 
$1/\kappa$ occurs for $|g|>1$ (panels (c) and (d)). In the latter situation the maximum 
diverges at a sufficiently large but finite bulk correlation length (panel (d)).}
\label{fig:1}
\end{figure}

For ionic impurities it is readily seen that $S_{11}(q)$ is monotonically decreasing with $q$ if
$|g|\leq1$ (see Figs.~\ref{fig:1}(a) and (b)).
If $|g|>1$ a maximum of $S_{11}(q)$ occurs at $q=q_\m{max}=\kappa\sqrt{|g|-1}$.
In this latter case $S_{11}(q_\m{max})=S_{11}(0)/(1-(\kappa\xi(|g|-1))^2)$ is finite for 
$\xi<(\kappa(|g|-1))^{-1}$ (see Fig.~\ref{fig:1}(c)), whereas $S_{11}(q)$ has a pole for 
$\xi\geq(\kappa(|g|-1))^{-1}$ (see Fig.~\ref{fig:1}(d)).
Obviously, whenever $|g|>1$, a divergence of $S_{11}(q)$ occurs upon approaching a critical point,
where the bulk correlation length $\xi$ diverges.
Hence, by adding ionic impurities the solvent remains uniform within the whole one-phase region of
the bulk phase diagram if and only if $|g|\leq1$, i.e., if the contrast of impurity-solvent
interactions is sufficiently small. 
This can be understood in such a way that a contrast of impurity-solvent interactions ($|g|>0$)
promotes density fluctuations of a particular finite wave length $2\pi/q_\m{max}$
which occur upon approaching a critical point.
The question on the origin of a sufficiently strong contrast of impurity-solvent interactions is
addressed in the following Subsec.~\ref{Subsec:origin}.

%---------------------------------------------------------------------------------------------------

\subsection{\label{Subsec:origin}Contrast of impurity-solvent interactions}

It has been shown in the previous Subsec.~\ref{Subsec:forms} that deviations from a monotonic
Ornstein-Zernike-like solvent structure factor $S_{11}(q)$ occur only for \emph{ionic} impurities
\emph{and} for a sufficiently \emph{large contrast} of impurity-solvent interactions ($|g|>1$).
Non-monotonic solvent structure factors have indeed been found in heavy water+3-methylpyridine 
mixtures in the presence of sodium tetraphenylborate ($\m{NaBPh_4}$) by means of small-angle
neutron scattering \cite{Sadakane2007b}.
However, the same solvent under the same conditions with simple salt impurities ($\m{LiCl}$, 
$\m{NaCl}$, $\m{NaBr}$, $\m{KCl}$ and $\m{MgSO_4}$) led to monotonic solvent structure
factors \cite{Sadakane2006,Sadakane2007a}.
This experimental observation implies that the valency of impurity ions are \emph{not} expected
to be the origin for impurity-induced non-uniformities of the solvent.
In agreement with this, it has recently been concluded within a theoretical model \cite{Bier2012} 
that neither the valency nor differences in solubility give rise to a sufficiently large parameter 
$|g|>1$.

Here it is proposed that $|g|>1$ may be reached due to steric effects, i.e., due to the property
of (ionic) impurities being structure-makers or structure-breakers.
For spherical particles $c_{12}^{(0)}\propto(R_1+R_2)^3$ and 
$c_{13}^{(0)}\propto(R_1+R_3)^3$ can be expected, where $R_i$ denotes the extension of a particle
of component $i$.
Hence, according to Eq.~\Eq{g2}, $|g|>1$ may be reached for impurities of significantly different
particle sizes such as $\m{Na^+}$ and $\m{[BPh_4]^-}$, whereas $|g|<1$ for particles of similar
size like simple salts. Non-monotonic solvent structure factors can be measured using small-angle 
neutron or x-ray scattering in view of the aforementioned typical Debye length $1/\kappa$ 
and $q_\m{max} / \kappa = \sqrt{|g|-1}$. In the case of highly diluted impurities the functional 
shape of the measured scattering intensity is dominated by the solvent structure factor.

It is important to understand that it is not the valency but the long-ranged
character ($\propto 1/r$) of electrostatic interactions of the impurities which gives rise to the
relevance of the parameter $|g|$ for the formation of a non-uniform solvent structure.
However, whether the solvent becomes non-uniform upon adding ionic impurities is mainly determined
by the size-differences between the impurity particles.

%===================================================================================================

\section{\label{Sec:conclude}Conclusions and Summary}

By means of general, model independent properties of the direct correlation functions it
has been shown that dilute impurities may influence the structure of a solvent if and only if they
are composed of ions with a sufficiently large contrast of impurity-solvent interactions.
It has been argued that the latter feature requires significantly different particle sizes.
Non-ionic impurities or ions of similar size are not expected to give rise
to a qualitative change in the large-scale structure of the solvent.
Ionic impurities with a sufficiently strong impurity-solvent interaction contrast lead to a 
non-monotonic solvent structure factor with a maximum at a magnitude of the scattering vector
close to the inverse Debye length (Fig.~\ref{fig:1}).
Upon approaching a critical point, this maximum height diverges at a finite bulk correlation 
length, which corresponds to the formation of a non-uniform bulk fluid.

For fluids with dilute impurities the asymptotic form of the structure factor derived here (see 
Eq.~\Eq{S11asym}) is expected to be a useful generalization of the standard Ornstein-Zernike form,
because additional information on the impurity-solvent interactions can be extracted (see 
Eq.~\Eq{g2}).
Application in future experimental studies on such systems would be highly welcome in order to
test the derived expression further. Finally, the Ornstein-Zernike approach can be applied to 
multicomponent impurities, i.e., $i \in \{1,2,3, ..., N\}$.
 
%===================================================================================================

\end{document}